\documentclass[floats,aps,prb,showpacs]{revtex4}
\usepackage{psfig}

\begin{document}

\title{High frequency oscillations in low-dimensional conductors
and semiconductor superlattices induced by current in stack
direction}

\author{S.N. Artemenko, S.V. Remizov}
\affiliation{Institute of Radio-engineering and Electronics of
Russian Academy of Sciences, 11-7 Mokhovaya str., Moscow 125009,
Russia}

\date{\today}
\begin{abstract}
Narrow energy band of the electronic spectrum in some direction in
low-dimensional crystals may lead to a negative differential
conductance and N-shaped I-V curve that results in an instability
of the uniform stationary state. Well-known stable solution for
such a system is a state with electric field domain. We have found
a uniform stable solution in the region of negative differential
conductance. This solution describes uniform high frequency
voltage oscillations. Frequency of the oscillation is determined
by antenna properties of the system. The results are applicable
also to semiconductor superlattices.
\end{abstract}


\maketitle

\section{Introduction}

There is a number of materials with narrow energy band of the
electronic spectrum for electrons moving in certain directions.
The narrow band may result either from low-dimensional structure
of the crystal or from artificially made periodic superstructure
like semiconductor superlattices. Typical examples of
low-dimensional crystals are layered materials such as high-Tc
superconductors, transition metal dichalcogenides like TaS$_2$,
NbSe$_2$ and others \cite{TMD}. Such materials can be treated as a
stack of conducting layers of atomic thickness, separated by
insulating ones \cite{Sloistost}. Narrow electronic bands are
typical also for quasi-one-dimensional conductors in directions
perpendicular to conducting chains. For example, the linear chain
conductor NbSe$_3$ is a strongly anisotropic crystal with the
anisotropy of conductivity $\sigma_b/\sigma_c \sim 10$ in (b-c)
plane, and $\sigma_b/\sigma_{a^*} \sim 10^4$ in (a-b) plane at low
temperatures\cite{Anisotropy}. Such a strong anisotropy reflects
weak inter-chain coupling and narrow electronic bands for electron
motion across conducting chains.

This feature of energy spectrum should lead to similarities of the
effects observed in low-dimensional materials and in semiconductor
superlattices which are thoroughly studied over last few decades
\cite{Wacker}. Interesting effects are expected if electron motion
approaches a regime close to the Bloch oscillations, when applied
voltage is high enough and an electron has enough time to perform
several oscillations within the same mini-band until it is
scattered, \textit{i. e.}, when period of the Bloch oscillations
is smaller then the momentum scattering time, $\tau_p$. In such a
regime DC current decreases with DC voltage increasing, resulting
in a region of negative differential conductance (NDC). This
result was predicted theoretically for superlattices by Esaki and
Tsu \cite{EsakiTsu}, and later was observed repeatedly. Similar
conclusions were made for layered conductors \cite{Artemenko}.
However, to our knowledge, there is no direct experimental
confirmation of this effect in low-dimensional conductors. A
possible reason for this is that the crystals were not perfect
enough to ensure the band motion of electrons in directions of low
conductivity in low-dimensional materials. However, the strong
conductivity maximum at zero bias voltage observed recently in
NbSe$_3$ by Latyshev \textit{et al.} \cite{Latyshev} argues in
favor of possibility of the NDC in such materials.

An important corollary of the NDC is an instability of uniform
stationary state in the NDC region which, in principle, may result
in generation of high frequency oscillations when the DC voltage
gets into the NDC region. The boundary of the NDC region in the
static I-V curve is determined by the voltage related to the
momentum scattering rate $\tau_p^{-1}$, and the upper frequency
limit for the NDC is equal to $\tau_p^{-1}$ as well. The value of
$\tau_p^{-1}$ in many materials falls into the THz region
\cite{Latyshev},\cite{TauP}, so one can expect microwave
generation in this frequency range. However, there is a number of
obstacles preventing generation of high frequency oscillations.
The primary one is formation of nonuniform electric field
distribution (field domains) which impedes high frequency
oscillations due to relatively low domain velocity \cite{Wacker}.
A number of techniques was suggested to overcome this difficulty
in superlattices, including application to a sample additional
large AC \cite{ACField} and using of pulse voltage \cite{RomanovImpulsy}, or
formation of superlattices with minibands having non-sinusoidal
dispersion law \cite{RomanovZona}, \textit{etc}.

In this work we study theoretically a low-dimensional conductor
with current flowing in the direction of low conductivity when
applied DC voltage  falls into the NDC region, and find a uniform
oscillatory solution in case when no external time-dependent
influence is imposed. This solution is possible because of
interaction of a sample with an environment which we describe by
means of an effective antenna impedance. We found conditions for
existence of stable oscillations: the antenna reactance on main
oscillation frequency must have the sign corresponding to an
inductance, in this case the sample and the effective antenna form
an effective oscillatory circuit. The frequency of the
oscillations is determined by the antenna reactance and by
material properties of the sample. Oscillation amplitude depends
on applied voltage and vanishes at the boundaries of the NDC
region.

The paper is organized as follows: in Sec.~\ref{secMainEq} we
derive general expression for the current across the layers.
In Sec.~\ref{secCoh} we study in details the uniform
current flow through the sample, derive equations of motion for
time-dependent voltage, and find a solution describing
oscillations. We show first the possibility of coherent voltage
oscillations using oversimplified but physically transparent
model, and then bring strict solution in the form of harmonic
series. In Sec.~\ref{secOtklik} we examine response of the system
to a small external time-dependent perturbation, and in
Sec.~\ref{secStab} we demonstrate that solution describing
coherent voltage oscillations in the sample is stable with respect
to small fluctuations. Finally, in Sec.~\ref{secDis} we discuss
conditions for experimental realization of the oscillatory regime
found in preceding sections and make conclusions.

\section{Main equations}
\label{secMainEq}

As it was mentioned in the Introduction, electron motion across
the layers of perfect enough layered conductor, or between the
conducting chains in quasi-one-dimensional conductor can
be considered as a motion in narrow conducting band that should
have specific features typical for electron motion within the
minibands in semiconductor superlattices. In the theory of
miniband transport there are two main limiting cases. If electron
wave functions in adjacent layers are strongly overlapped then an
electron has enough time to perform several tunneling processes
until it is scattered. This quasi-classical case of miniband
conduction can be well described by means of Bolzman equation
which holds when $\tau_p^{-1} \ll \Delta$, where $\Delta$ is the
miniband width. If the interlayer hopping integral is small then
the sample may be treated as a stack of weakly coupled quantum
wells, hence, tunneling processes in adjacent junctions are
practically independent, and purely quantum case of sequential
tunneling is fulfilled. An amazing result of supperlattice theory
is that in both these limits expressions describing the current as
functional of voltage are the same \cite{Wacker}. Below we show
how the similar expression can be derived for a low-dimensional
conductors. For definiteness we will speak about a layered
conductor with voltage applied in the stack direction.

For brevity we use units with $e= \hbar=1$. To investigate
non-equilibrium transport we use the Keldysh technique for Green
functions \cite{Landau}. Current density between layers $n$ and
$n+1$ may be expressed in terms of the non-diagonal component of
Green's function both on temporal indices and layer numbers:
$j_{n, n+1}(\tau) = i e t_{\perp}\hbar^{-1} \left(G^K_{n, n+1}(\tau, \tau) -
G^K_{n+1, n}(\tau, \tau)\right)$,
$$
  \hat G_{n, m} = \left(
    \begin{array}{cc}
      G^R_{n, m} & G^K_{n, m}\\
      0&G^A_{n, m}
    \end{array}
  \right)
,
$$
where $t_{\perp}$ is the hopping integral for electron tunneling between the adjacent layers, which is supposed to be independent on layer number and electron momentum, and $G^{R(A)}$ is the retarded (advanced) Green function, $\tau$ is the dimensionless time, which is mesured in units of $\tau_p$. For convenience we introduce other dimensionless units: frequencies and energies we mesure in terms of $\tau_p^{-1}$. 

Usually, the hopping integral, $t_{\perp}$, is small and it is
possible to construct a perturbation theory, which allows us to
express non-diagonal on layer number Green functions in terms of
diagonal ones. In the first approximation on $t_{\perp}$ it reads
$\hat G_{n, n+1}(\tau, \tau') = t_{\perp} \int \hat G_{n, n}(\tau, \tau'')
\hat G_{n+1, n+1}(\tau'', \tau') d\tau''$, and, hence, we get well known
expression for the current between the adjacent layers
\cite{GreenFunction}.

$$
  j_{n, n+1}(\tau) = i t_{\perp}^2 \int G^R_{n, n}(\tau, \tau') G^K_{n+1, n+1}(\tau',\tau)
   + G^K_{n, n} (\tau, \tau') G^A_{n+1, n+1}(\tau', \tau) d  \tau'
.
$$
We consider the case when electrons inside the layer are in in the
equilibrium while layers are not in the equilibrium with each
other. A difference between $j_{n-1, n}$ and $j_{n,
n+1}$ leads to deviation $\delta \mu_n$ of the chemical potential
in layer $n$ from its equilibrium value, $\mu$. In this case in each layer the diagonal Green function with respect to
layer number, $G^K_{n, n}$, can be related to the retarded and
advanced Green functions by the expression valid for the
equilibrium case: $
  G^K_{n, n}(\omega) =
    (G^R_{n, n}(\omega) - G^A_{n, n}(\omega))
    \mathop{\rm tanh} (\omega - \delta \mu_n)/2 T
, $ 
here we take into account temporal uniformity, so that $G^{R(A)}_{n, n}(\omega, \omega') = G^{R(A)}_{n, n}(\omega) \delta(\omega - \omega')$.
The transition to temporal representation
we perform according to
$$
  G^{R(A)}_{n, n}(\tau', \tau'') =
  \int e^{i \omega' \tau' - i \omega'' \tau''} G_{n, n}^{R(A)}
  (\omega', \omega'') e^{\int_{\tau'}^{\tau''} \Phi_n(\tau) d\tau} d\omega' d\omega'',
$$
here $\Phi_n(\tau)$ is time-dependent scalar electric potential of
corresponding layer. 
To give proper account to scattering, we use the standard expression for Green functions with non-zero momentum relaxation
rate $
  G^{R(A)}_{n, n}(\omega)=(\omega - \xi_p \pm i/2)^{-1}
, $ here ``$\pm$'' is related to the retarded (advanced) Green
function, respectively, $\xi_p$ is the electron energy, counted
from the Fermi energy, $p$ is the component of electron momentum
parallel to the layers. 

We assume for simplicity that $\tau_p$ and $t_\perp$ do not depend
on electron momentum. We concentrate on the case of the same voltage on all junctions, taking the non-uniformity into account only in
Sec.~\ref{secStab} where we study the stability of the uniform solution in
respect to small perturbations. So we assume that deviations,
$\delta \mu_n$, of the chemical potential from the equilibrium
value are small.

Finally, the
expression for the tunneling current between adjacent layers reads
\begin{eqnarray}
\nonumber
  j_n^{(t)}(\tau) &=& \frac{\sigma_0}{s}\left(
    \int \limits_0^{\infty} d \tau_1 e^{-\tau_1} \sin \int \limits_0^{\tau_1} \left(
      V_n(\tau) - V_n(\tau - \tau_2)
    \right) d\tau_2
    -
  \right.
\\
  &&\left.
    (\delta \mu_{n+1} - \delta  \mu_n) \int \limits_0^{\infty} d \tau_1 e^{-\tau_1}
    \cos \int \limits_0^{\tau_1} \left(
      V_n(\tau) - V_n(\tau - \tau_2)
    \right) d\tau_2
  \right)
, \label{j1pro}
\end{eqnarray}
where $s$ is the lattice period in the direction of the current.
The second item in square brackets in Eq. (\ref{j1pro}) describes
the diffusion current, $V_n(\tau)$ is the voltage between
layers $n+1$ and $n$, $\sigma_0=m e^2 t^2_{\perp}\tau_p/\hbar $ is
conductivity at zero bias and frequency, and $s$ is the period of
the structure.

Consider now limiting forms and simplifications of equation
(\ref{j1pro}) for the current. We represent the voltage across the
layer as a sum $V(\tau)=v_0 + v(\tau)$, where $v_0$ is the average
voltage (here we drop index $n$). When the second time-dependent
term is small Eq.~(\ref{j1pro}) can be expanded in series with
respect to $v(\tau)$, and we obtain for the current:
\begin{equation}
 j^{(t)}(\tau) = j_0 + \frac{1}{s}\left(
  \sigma_d v(\tau) +
  \frac{1}{2}\sigma'_d v(\tau)^2
    + \frac{1}{6}\sigma''_d v(\tau)^3
  \right)
   + (c_t + c'_t v(\tau))
   \partial_\tau v(\tau)
. \label{jSimpl}
\end{equation}
Here we introduce the DC part of the current, $j_0 (v) = \sigma
v_0/s$ where $\sigma = \sigma_0 /(1+v_0^2)$. Further, $\sigma_d
\equiv j'_0 \equiv dj_0/dv_0  = \sigma_0 (1-v_0^2)/(1+v_0^2)^2$ is
the differential conductance, and $c_t = (3 v_0^2-1)/s(1+v_0^2)^3$
is the specific tunneling capacity of the sample. The apostrophe
denotes derivative on voltage $v_0$.

As an illustration, consider, first, the simplest case when the
applied voltage is time-independent, $v(\tau) = 0$. Then
Eq.~(\ref{jSimpl}) results in well-known form \cite{EsakiTsu}: $j
= j_0 \sim v_0 (1+v_0^2)^{-1}$. It follows from this expression
that the differential conductance becomes negative when $v_0 > 1$.

In case of small harmonic voltage, $v(\tau) = v_1 e^{i
\omega \tau}$ the expression for the current in the first
approximation in $v(\tau)$ reads
\begin{equation}
j^{(t)}(\tau) =  j_0 +
  v_1 \frac{i \omega (v_0^2-1-i\omega) \sigma_0}{s \left(v_0^2-(\omega-i)^2\right)\left(v_0^2+1\right)} e^{i \omega \tau}
. \label{eqV1pV2w}
\end{equation}
From this expression it follows that at high frequencies, $\omega
\gtrsim 1$, the differential conductivity becomes positive even
when the DC bias is in the NDC region.

\section{Coherent voltage oscillations}
\label{secCoh}

In this section we derive equation of motion for the voltage and
study dynamics of the system when currents in all junctions are
identical. In addition, we neglect voltage fluctuations here,
therefore, we do not consider non-equilibrium deviations of
chemical potential in different layers and, hence, the diffusion
current.

In addition to the tunneling current (\ref{j1pro}) there are other
contributions to the total current. First, we must take into
account the displacement current
\begin{equation}
  j^{(displ)} =  c_0 \frac{\partial V(\tau)}{\partial \tau}
, \label{j}
\end{equation}
where we use notation for the specific capacity of the sample $c_0
= 1 / 4 \pi s$.

We must take into account also leakage currents, contributions due
to incoherent tunneling, and other processes which are not related
to coherent tunneling. To allow for these effects
phenomenologically we add an extra ohmic term to the current:
\begin{equation}
  j^{(ohm)}(\tau)=\sigma_1 \frac{V(\tau)}{s}
. \label{j2}
\end{equation}
For the I-V curve to have the NDC region conductivity $\sigma_1$
must be small enough:
\begin{equation}
  \sigma_1 < \sigma_0 / 8
. \label{SigmaCond}
\end{equation}

\begin{figure}[!ht]
  \vskip 0mm
  \centerline{
    \psfig{figure=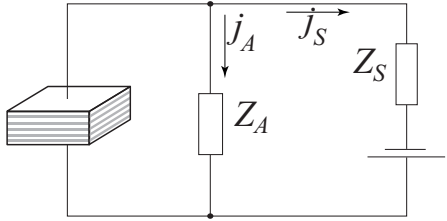,height=3cm
    ,angle=0}
  }
  \caption{
    Equivalent circuit of the system. $Z_A$ and $Z_S$ are impedances of the antenna
    and of the power supply, $j$, $j_A$, and $j_S$ are the currents flowing through the
    sample, antenna, and power supply, respectively. Current in the external
    circuit $j^{(ext)} = j_S + j_A$. We assume  on frequencies of the problem
    $|Z_A| \ll |Z_S|$ and on zero frequency $|Z_A| \gg |Z_S|$ so that $j_A$ is AC and $j_S$ is
    DC.
  }
\label{figScheme}
\end{figure}

To determine two independent variables, the current and the
applied voltage, an additional equation relating the applied
voltage to the current in the external to the sample circuit is
required. The equation for current balance which determines the voltage dynamics reads
\begin{equation}
  j^{(t)} + j^{(ohm)} + j^{(displ)} + j^{(ext)}=0
\label{main}
\end{equation}
Here $j^{(ext)} = j_A + j_S$ is the current in the circuit external to the sample. We consider this circuit as that consisting of a
power supply and environment. Coupling of the sample to the
environment we describe by means of an effective antenna with the
impedance $Z_A$. The equivalent scheme for the system under consideration is shown in Fig.~\ref{figScheme}. Since typical frequencies of the problem is high, we assume that AC current does not flow through power supply, so that antenna current $j_A$ is AC and power supply current $j_S$ is DC.

Sample dimensions are assumed to be small with respect to radiation wavelength and to the skin-effect length. This allows us to neglect spatial non-uniformity.

The resulting equation describing dynamics of the voltage
oscillations has rather complicated form of the
integro-differential equation. In order to find an analytical
solution we consider the limit in which the amplitude of the
voltage oscillations is small in comparison to the DC voltage,
$v(\tau) \ll v_0$. This regime occurs near the boundary of the NDC
region or in the case when the NDC region is small.

\subsection{Simple  model}

It is instructive to examine, first, an oversimplified, but
physically transparent model, based on assumption that in the
frequency range under consideration the effective antenna can be
described by the admittance with purely inductive reactive part:
$j_A(v) \to \frac{N}{LS}\int v(\tau) d\tau $. Here $L$ is the
antenna inductance, and factor $N$ is number of layers in the
sample, so that the total voltage across the sample is equal to
$N v(\tau)$. The real part of the antenna admittance leads to
redefinition of the conductivity $\sigma_1$ only, therefore, we
drop it here. For convenience we introduce here a new variable related to
the uniform alternating electric field in the sample as $\dot \theta = v(\tau) / s$, and assume that the NDC region is
small, so that contribution of the second harmonic into the
friction terms is negligible (see the next subsection).
Eq.~(\ref{main}) then reads:
\begin{equation}
  \ddot \theta + \omega_0^2 \theta + \frac{1}{c_s s} \left(
    \delta \, \dot \theta +
    \frac{\sigma'_d \dot \theta^2}{2} +
    \frac{\sigma''_d \dot \theta^3}{6} -
    j +       \sigma_1  \frac{v_0}{s}
  \right)
    = 0
. \label{equ}
\end{equation}
Here $j$ is DC current flowing through the sample, $\omega_0 = \sqrt{N / L S c_s }$, $S$ is the sample area in
the plane perpendicular to the current direction, and $N$ is the
number of junctions, and we have introduced the total specific
capacity of the sample, $c_s = c_0+c_t$, and supercriticality,
$\delta = \sigma_d + \sigma_1$, which changes its sign from
``$-$'' to ``$+$'' across zero when the DC voltage leaves the
instability region of the I-V curve.

Equation for oscillations (\ref{equ}) can be treated as a
mechanical equation of motion. In this equation $\theta$ plays a
role of a coordinate, the last term in the left-hand side of
Eq.~(\ref{equ}) corresponds to a time-dependent alternating
friction. In the regime of stable oscillations the friction
averaged over the period must be equal to zero, and its momentary
value can be considered perturbatively. So, in the first
approximation we can neglect the friction term, $
  \ddot \theta^{(1)} + \omega_0^2 \theta^{(1)} = 0
. $ Since the phase of the solution is arbitrary, for
definiteness, we choose it in the form $
  \dot \theta^{(1)} = E_1 \sin \omega_0 \tau
$. $E_1$ is an amplitude of electric field oscillations on main frequency, so voltage oscillations amplitude on this frequency is $E_1 s$. To ensure that the solution does not increase with time the
friction term in (\ref{equ}) must not contribute at the
oscillation frequency $\omega_0$. This condition determines the
amplitude of the oscillations
\begin{equation}
  E_1 =
\sqrt{- \frac{8 \delta}{\sigma''_d}} .
\label{solA1Eq}
\end{equation}

\begin{figure}[!ht]
  \vskip 0mm
  \centerline{
    \psfig{figure=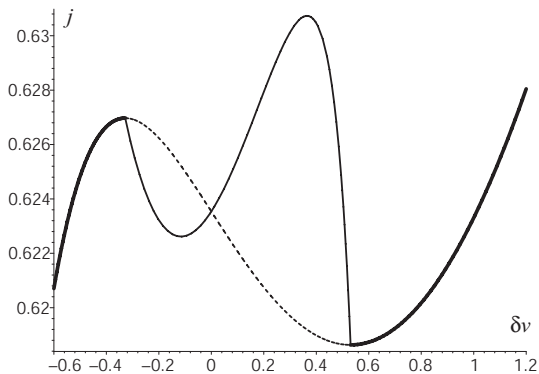,height=5cm
    ,angle=270}
  }
  \caption{
    Dynamic I-V curve is shown with the bold line. The dashed line describes the static I-V curve.
    Current is measured in dimensionless units, voltage $\delta v$ is counted from $v_0 = \sqrt{3}$
    corresponding to the middle of a small NDC region. The I-V curve is presented for
    $\sigma_1=0.11$.
  }
\label{figVAH}
\end{figure}

Guided by mechanical analogy mentioned above, we multiply friction
term in Eq.~(\ref{equ}) by $\theta_1$ and integrate over the
period and demand that the work of the friction force over the
oscillation period is zero. The condition of vanishing of the
average friction determines the I-V curve:
\begin{equation}
  j(v_0) =
    j_{stat} -
    2\frac{\sigma'_d\delta}{\sigma''_d}
. \label{VAHd}
\end{equation}
This dynamic I-V curve intersect the static one $j_{stat} =
\left(\sigma_1+\sigma_0/(1+v_0^2)\right) v_0$ at three points: at the
boundaries of the NDC region and at $v_0=\sqrt{3}$ (see
fig.~\ref{figVAH}).

\subsection{Solution in the form of harmonic series}

In this subsection we consider a more general case and determine
conditions for existence of the oscillatory solution. We seek the
solution of Eq.~(\ref{main}) for $v(\tau)$ in the form of harmonic
series
\begin{equation}
  v(\tau)=\sum\limits_{n=1}^\infty(v_{s,n} \sin(n \omega_0 \tau) +
  v_{c, n} \cos(n \omega_0 \tau))
, \label{sol}
\end{equation}

In the most general case the antenna current reads
$$
  j_A=
  \frac{N}{s}\sum\limits_{n=1}^{\infty}(\sigma^{R}_{A n} v_{s, n}+\sigma^{I}_{A n} v_{c, n})
  \sin n \omega \tau + (\sigma^{R}_{A n} v_{c, n} -
  \sigma^{I}_{A n} v_{s, n}) \cos n \omega \tau
.
$$
Here $\sigma^{R}_{A n}=R_n s/(R_n^2+X_n^2)S$, $\sigma^{I}_{A
n}=X_n s/(R_n^2+X_n^2)S$, where $R_n$ and  $X_n$ are resistance
and reactance of the antenna at frequency $n \omega_0$ respectively.

Solution (\ref{sol}) has an arbitrary phase which can be
determined by initial conditions. We may fix this phase without
any restriction assuming that $v_{c, 1} = 0$. The supercriticality
parameter, $\delta_n = N \sigma^{R}_{A n} + \sigma_1 + \sigma_d$, we
assume to be small at the main oscillation frequency: $|\delta_1|
\ll 1$. As it will be shown later, amplitudes of the $n$-th
harmonic are proportional to $\delta_1^{n/2}$, hence, it is
possible to neglect the terms of the order higher than the third
one in the small parameter $\sqrt{\delta_1}$. So inserting Eq.
(\ref{sol}) into the expression for the current (\ref{j1pro}), and
using equation for the current balance (\ref{main}), we obtain a
system of linear equations for $v_{s, 1}^2$, $v_{s, 2}$ and $v_{c,
2}$:
\begin{equation}
\left\{
\begin{array}{rcl} 
-\sigma''_d v_{s, 1}^2
+ 4 \omega c'_t s v_{s, 2}
+ 4 \sigma'_d v_{c,2}
&=&
8 \delta_{1} s
\\ 
\sigma'_d v_{s, 2}
-c'_t s \omega v_{c, 2}
&=&
2 N \sigma^{I}_{A1} - 2 c_s s \omega 
\\ 
\omega c'_t s v_{s, 1}^2
+ 2 \delta_{2} v_{s, 2}
-2 \sigma_2 v_{c,2}
&=&
0
\\ 
\sigma'_d v_{s, 1}^2
- 4 \sigma_2 v_{s, 2}
- 4 \delta_{2} v_{c,2}
&=&
0.
\end{array}
\right. \label{matrixA}
\end{equation}
Here $\sigma_2 = 2 c_s \omega - N \sigma^{I}_{A 2}$.
Physically reasonable oscillation frequency is determined by the second equation of the system
(\ref{matrixA}): 
\begin{equation}
  \omega \simeq \frac{N \sigma^{I}_{A 1}}{c_s s}
. \label{solOmegaSeries}
\end{equation}
Here we take into account that the contribution of the second harmonic is of the second order in $\sqrt{\delta_1}$,  so that left-hand sinde of this equation should be considered as higher order correction and, hence, should be neglected in main approximation. 
For oscillation amplitudes we find
\begin{eqnarray}
  v_{s, 1}^2 &=&
    - 8 \delta_1 \frac
    {{\sigma_2} ^2 + \delta_{2}^2}
    {\mathcal{D}}
\label{vs12Full}
\\ v_{s, 2} &=& 2\delta_{1} \frac{2 \omega \delta_{2} c'_t s- \sigma'_d \sigma_2}{\mathcal{D}}
\\ v_{c, 2} &=& - 2\delta_{1} \frac{2 \omega c'_t s\sigma_2+\delta_{2} \sigma'_d}{\mathcal{D}}
,
\end{eqnarray}
\begin{equation}
  \mathcal{D}=-3 \sigma'_d \omega c'_t s\sigma_2 + (2 \omega^2 {c'_t}^2 s^2 - {\sigma'_d}^2)\delta_2
  +
  (\sigma_2^2 +  \delta_{2}^2) \sigma''_d
.\label{eqD}
\end{equation}
This expressions have a transparent physical meaning: if $\mathcal{D}>0$ the
existence of the oscillatory regime is determined by the sign of
the supercriticality, $\delta_1$. Oscillations are possible,
({\textit i.e.}, the amplitude of the first harmonic is real) only
inside of the NDC region.

Now we show that $\mathcal{D}$ is positive under conditions of our assumption
that harmonics drop rapidly. If the NDC region is small then the
voltage is close to the center of the NDC region, $|v_0 -
\sqrt{3}| \ll 1$. Then $\sigma'_d \propto v_0^2 - 3$ is small and
$\mathcal{D}$ is positive. If the NDC region is large then our approach is
valid near the boundaries of the NDC region. Near the low-voltage
boundary $v_0 \sim 1$. In this case $\sigma_d$, $c'_t \to 0$ as
$v_0-1$ and $\sigma'_d \simeq 1/2$, $\sigma''_d \simeq 3/2$. So if
$\delta_2$ and/or $\gamma_2$ are large enough, then $\mathcal{D}>0$. The
high-voltage boundary of the NDC region corresponds to $v \gg 1$.
In this case $c'_t$, $\sigma'_d \sim v_0^{-3}$, $\sigma''_d \sim
v_0^{-4}$, and two first terms in (\ref{eqD}) are small in
comparison to the last one and, hence, $\mathcal{D}>0$. In other words, in
this case the influence of second harmonic on the amplitude of the
first one is small, so that $v_{s,1} \simeq \sqrt{-8
\delta_1/\sigma''_d}$, as it was found in the previous subsection.

At the end of this chapter we consider the amplitudes of the
higher harmonics. Acting perturbatively, we can express the
amplitude of the third harmonic in terms of amplitudes of the
first and the second harmonics considering the last ones as an
``external force'':
\begin{eqnarray}
  v_{c,3} = -v_{s,1} \frac
    {-{\sigma^{I}_{A3}} \sigma''_d v_{s,1}^2+12 \sigma'_d ({\sigma^{I}_{A3}} v_{c,2} - \delta_{3} v_{s,2})}
    {24 (\delta_{3}^2+{\sigma^{I}_{A3}}^2)} \sim \delta_1^{3/2},
\\
  v_{s,3} = - v_{s,1} \frac
    {12 \sigma'_d  ({\sigma^{I}_{A3}} v_{s,2} + \delta_{3} v_{c,2}) - \delta_{3} \sigma''_d v_{s,1}^2}
    {24 (\delta_{3}^2+{\sigma^{I}_{A3}}^2)} \sim \delta_1^{3/2}.
\end{eqnarray}

Amplitudes of the higher harmonics, $n>3$, can be found in the
similar perturbative way, and one can see that their amplitudes
are proportional to $\delta_1^{n/2}$. Therefore, the harmonic
series (\ref{sol}) converges and our perturbative approach is
valid provided $|\delta_1| \ll 1$.

Thus in this section we have found the solution describing voltage
oscillations in the vicinity of NDC region boundaries or for whole
NDC region if it is small. Beyond the NDC region boundaries the
oscillatory solution does not exist, this is reflected by the
imaginary value of the oscillation amplitude in
Eq.~(\ref{vs12Full}). Amplitude of the oscillations vanishes when
the average voltage approaches NDC region boundary. Oscillation
frequency is determined by geometry and by material properties of
the sample. This solution exists provided the imaginary part of
impedance of the effective antenna has the sign corresponding to
inductance.

\section{Microwave Response}
\label{secOtklik}

In this section we determine modification of the I-V curve when
the system is exposed to a weak external harmonic radiation of frequency
$\omega$. Again, for simplicity, we treat antenna as an
inductance. Let the incident radiation induces voltage
oscillations with the amplitude $V_A$ at the antenna output. Then
the equation for the oscillating component of the voltage
(\ref{equ}) acquires the form
\begin{equation}
  \ddot \theta + \omega_0^2 \theta +
  \frac{\delta \dot \theta +
    \frac{1}{2} \sigma'_d \dot \theta^2 +
    \frac{1}{6} \sigma''_d \dot \theta^3
  }{c_s s}
  =
  \frac{j - (\sigma_1+\sigma_2)v_0 /s}{c_s s} + f \sin \omega \tau
. \label{RadiatEq}
\end{equation}
Here $f = V_A \omega_0^2 /\omega N s$ is the amplitude of the external force. This
equation describes the forced oscillations of the system with
single degree of freedom in case of periodic perturbation. As in
the previous section we consider the case when the amplitude of
the voltage oscillations is small and we solve equation for
$\theta(\tau)$ perturbatively. When the frequency of the external
field, $\omega$, is close to the eigen frequency, $\omega_0$, one
can expect the effect of frequency capture. As in the previous
section we consider the case when the amplitude of the voltage
oscillations, $a$, is small. Then in the first approximation we
seek the solution in the form of harmonic oscillations at the
frequency of the external force but with unknown time-independent
phase shift $\psi$: $
  \theta(\tau) = a \sin (\omega \tau + \psi)
$. Then we find from Eq.~(\ref{RadiatEq})
\begin{equation}
  a\sqrt{A_s^2 + A_c^2} \sin(\omega \tau + \psi + \varphi)
  =
  f \sin(\omega \tau)
, \label{RadiatEq1}
\end{equation}
where $A_s = \omega_0^2-\omega^2$, $A_c = \omega ( \delta + a^2
\omega^2 \sigma_d''/8) / c_s s$, and $\tan \varphi = A_c/A_s$.
Eq.~(\ref{RadiatEq1}) determines the amplitude of oscillations as
function of frequency, $\omega$, and bias voltage, $v_0$. We seek
solution of Eq.~(\ref{RadiatEq1}) for the amplitude in the form
$$a= \frac{E_1}{\omega}(1+y), $$
where the first term describes the voltage oscillations with the
amplitude of the natural oscillations, $A_1$, given by
Eq.~(\ref{solA1Eq}). For small enough values of $f$ we find two
solutions describing corrections to the oscillation amplitude
induced by the external perturbation
\begin{equation}
  y (\omega, v_0) = \pm \frac{c_s s}{2 \omega \delta}
  \sqrt{\left(\frac{f\omega}{E_1}\right)^2
  -(\omega^2-\omega_0^2)^2}.
\label{RadiatHiDiss}
\end{equation}
This expression is valid provided the external perturbation is
small enough, $|f| \ll \delta^{3/2}$.

Corresponding expressions for the phase read
\begin{equation}
 \psi = \pm \arctan\frac{\sqrt{(f\omega/E_1)^2
  -(\omega^2-\omega_0^2)^2}}{|\omega^2-\omega_0^2|}.
 \label{RadiatPhiDiss}
\end{equation}
Using the approach of the next section one can show that both
solutions found above are stable.

For $\omega = \omega_0$ the phases of the solutions are equal to
$\pm \pi/2$. In this case the first solution describes voltage
oscillations in phase with the driving voltage, the oscillation
amplitude being increased by the external perturbation, while the
second solution corresponds to the out of phase oscillations with
the amplitude decreased by the external force.

According to Eq.~(\ref{RadiatHiDiss}) the regime of frequency
capture exists in the frequency range
\begin{equation}
 |\omega^2-\omega_0^2| < \frac{f\omega}{E_1}.
 \label{Freqrange}
\end{equation}
Variation of the I-V curve due to irradiation as function of
frequency and DC bias is described by expression
$$
\Delta j = \frac{\sigma '_d \omega^2 E_1^2}{2}y.
$$
Thus we can conclude that the external radiation induces two branches at the I-V curve that merge near the boundary of the regime of the frequency capture. Maximum variation of the amplitude and, hence, the largest shift of the I-V curve occur at $\omega = \omega_0$. They are linear in driving voltage.

At larger difference between the driving frequency and the eigen frequency, that is outside the regime of the frequency capture, corrections to the I-V curve induced by the external perturbation are quadratic in $f$.

\section{Stability}
\label{secStab}

Now we study the stability of the solution found in preceding section with respect to non-uniform perturbations of the voltage. Consider a sample consisting of $N$ layers. As it was stated above, we assume non-equilibrium processes to be slow enough in the scale of the energy relaxation time, so that distribution function of electrons inside conducting layers does not change. Then the in-layer electron distribution can be described by the Fermi distribution function with energy shifted by a non-equilibrium correction to the chemical potential induced by difference of the currents through adjacent conducting layers. At first we relate the shift of the chemical potential to the voltages in adjacent layers. Taking into account quasi-2D character of the layers, we obtain the expression for the difference of chemical potentials on the adjacent layers
\begin{equation}
  \delta \mu_n(\tau) = \frac{\hbar^2}{4 m e s}(V_n(\tau) - V_{n-1}(\tau))
. \label{mu}
\end{equation}
This allows us to express current in terms of voltages across
different junctions only:
\begin{eqnarray}
\nonumber
  j_n^{(t)}(\tau) &=& \frac{\sigma_0}{s}\left[
    \int \limits_0^{\infty} d \tau_1 e^{-\tau_1} \sin \int \limits_0^{\tau1} \left(
      V_n(\tau) - V_n(\tau - \tau_2)
    \right) d\tau_2
  \right.
\\
  &&
  \left.
    -
    \frac{\partial_n^2 V_n(\tau)}{4 m s} \int \limits_0^{\infty} d \tau_1 e^{-\tau_1}
    \cos \int \limits_0^{\tau_1} \left(
      V_n(\tau) - V_n(\tau - \tau_2)
    \right) d\tau_2
  \right]
, \label{j1}
\end{eqnarray}
where $\partial_n^2 V_n = V_{n+1} + V_{n-1} - 2 V_n$ is a discrete
version of the second derivative on layer number.

Since sample dimensions are assumed to be small, the current in
the external circuit is determined only by the total voltage
across the sample:
\begin{equation}
  j^{(ext)}(\tau) = \hat \sigma_A \left[
    \frac{1}{sN} \sum \limits_n V_n(\tau)
  \right]
. \label{jext}
\end{equation}
Using Eqs.~(\ref{j1}) and (\ref{jext}) in Eq.~(\ref{main}) we
obtain the equation describing voltage dynamics taking into
account dependence of the voltage on the layer number:
\begin{equation}
    j_n^{(t)} + \sigma_1 V_n(\tau) + \frac{1}{4 \pi s} \frac{\partial}{\partial \tau} V_n(\tau)
  =
  \hat \sigma_A \left[\frac{1}{sN} \sum \limits_k V_k(\tau)\right]
. \label{main1}
\end{equation}
We simplify Eq.~(\ref{main1}) assuming that in the considered
frequency range the antenna impedance can be modeled by pure
inductance,
$$
  \hat \sigma_A \left[\frac{1}{sN} \sum \limits_k V_k(\tau)\right] \to \frac{1}{LS}\int \sum \limits_k v_k(\tau) d\tau
.
$$
Then each equation in Eq.~(\ref{main1}) acquire a form of
Eq.~(\ref{equ})
\begin{equation}
  \ddot \theta_n + \frac{\omega_0^2}{N} \sum_k \theta_k + \frac{1}{c_s s} \left(
    \delta \dot \theta_n +
    \frac{\sigma'_d \dot \theta_n^2}{2} +
    \frac{\sigma''_d \dot \theta_n^3}{6} -
    j +
    \left(
      \sigma_1 + N \sigma_A
    \right) \frac{v_0}{s}
  \right)
    = 0
. \label{equs}
\end{equation}
To show that solution describing coherent voltage oscillations is
stable, we substitute $\theta_n(\tau)=\theta^{(0)}(\tau) +
\delta_n(\tau)$ into (\ref{equs}), where $\theta^{(0)}(\tau)$ is
the solution found in Sec.~\ref{secCoh}, and then linearize the
resulting equation in $\delta_n$. Temporal evolution of small
perturbation $\delta_n(\tau)$ is described by equation:
\begin{equation}
  \ddot \delta_n +
  \left(
    \sigma(\tau)
    - D \partial_n^2
  \right)
  \dot \delta_n +
  \frac{\omega_0^2}{N} \sum_k \delta_k = 0
\label{eqdelta} .
\end{equation}
Here $\sigma(\tau) =
  \left(
    E_1
    \sigma_d
    \cos(\omega_0 \tau + \varphi_0)+
    \frac{1}{8} \sigma''_d E_1^2 (1 + 2 \cos 2 (\omega_0 \tau + \varphi_0)
  \right)/c_s
$ plays the role of time-dependent friction coefficient, and $D =
\sigma_0 /4 m c_s s^2 (1+v^2_0)$ is diffusion coefficient.

The general solution of linear equations (\ref{eqdelta}) can be
divided into a uniform and a non-uniform parts: $\delta_n =
\delta^{(0)}+\delta^{(n)}$, $n=1..N$. The uniform part of the
solution, $\delta^{(0)}$, can be found as the general solution of
the equation
\begin{equation}
  \ddot \delta^{(0)}(\tau) + \sigma(\tau) \dot \delta^{(0)}(\tau) + \omega_0^2\delta^{(0)}(\tau) = 0
, \label{eqdeltaos}
\end{equation}
which follows from Eq.~(\ref{eqdelta}) when all $\delta_n$ are
equal. The rest of Eq.~(\ref{eqdelta}) form $N$
identical equations for the non-uniform part of the perturbation, $\delta^{(n)}$
\begin{equation}
  \ddot \delta^{(n)}(\tau) + [\sigma(\tau) - D \partial_n^2] \dot \delta^{(n)}(\tau) = 0
, \label{eqdeltan}
\end{equation}
which satisfy the condition which ensures that the total voltage
over all layers is described by $\delta^{(0)}$
\begin{equation}
  \sum\limits_{n=1}^N \delta^{(n)}(\tau) = 0
.
\label{eqPertCond}
\end{equation}

So we examine the temporal evolution of the total voltage across
the sample and voltage fluctuations at different junctions
separately, by Eqs.~ (\ref{eqdeltaos}) and (\ref{eqdeltan}),
respectively.

At first we study non-uniform perturbations described by
Eq.~(\ref{eqdeltan}). Its solution can be found easily by means of
the Fourier transformation on layer number:
\begin{equation}
  \delta_k(\tau) = c_1(k) + c_2(k) \int e^{-\int (\sigma(\tau) + D \hat k^2) d\tau} d\tau,
  \label{deltasol}
\end{equation}
where $\hat k = 2 \sin k/2$, and $c_1(k)$ and $c_2(k)$ are
constants of integration. Constants $c_1(k)$ are irrelevant since
they are not related to voltage fluctuations since voltage is
determined by time derivative of the phase $\theta_n$, so we can
equate these constants to zero. Since the diffusive term promotes
the stability ($D>0$), we concentrate on the case of small $\hat k
\to 0$. Then the stability of the solution is determined by the
damping term $\sigma(\tau)$ which can be presented as
$\sigma(\tau) = \sigma_0+\sigma_1(\tau)$ where $\sigma_1(\tau)$ is
a sum of harmonic terms, and the average value of the damping,
$\sigma_0$, is positive. Since it is positive, Fourier transforms
of the perturbation (\ref{deltasol}) decrease exponentially at
$\tau \to \infty$ for all values of $k$.

Note that the results on stability with respect to non-uniform
oscillations remain valid in the general case, when the external
antenna cannot be described by pure inductance. Indeed, in case of
antenna described by a linear operator of the total voltage across
the sample, the uniform and non-uniform fluctuations can be
examined separately similar to the way it was described above.
Thus non-uniform perturbations cannot lead to instability of the
oscillatory solution and the stability of the oscillatory regime
is determined only by the stability of uniform oscillations.

Now we discuss the uniform perturbation. The general solution of (\ref{eqdeltaos}) is
\begin{equation}
  \delta^{(0)}(\tau) =
  C_1 v(\tau) +
  C_2 v(\tau) \int \frac{1}{v^2(\tau)} e^{- \int \sigma(\tau) d\tau} d\tau
. \label{solDelta}
\end{equation}
Here $C_1$ and $C_2$ are arbitrary constants. As, by definition,
$\dot \theta^{(0)}(\tau) = v(\tau)$, the first term in
(\ref{solDelta}) describes the solution equal to the unperturbed
solution that is shifted by time $C_1$. This term is related to
the different choice of initial conditions, and it is not related
to stability. The second term decreases exponentially. This can be
shown by direct calculations and demonstrated by means of the
Ostrogradsky--Liouville formula relating the Wronskian to the
coefficients of the linear differential equation,
$$
\begin{array}{|cc|}
  v(\tau) & \delta(\tau)
  \\
  \dot v(\tau) & \dot \delta(\tau)
\end{array}
=C e^{- \int \sigma(\tau)  d\tau} ,
$$
where $C$ is an arbitrary constant. With $\tau$ increase perturbed
solution exponentially converges to non-perturbed one which may
have some phase shift with respect to original solution, {\it
i.e.}, original uniform solution is stable.

\section{Discussion}\label{secDis}

In this section we discuss some aspects of experimental
realization of the oscillatory regime. The condition for voltage
oscillation derived in the previous section implies that the
supercriticality must be negative, $\delta_1<0$. This is possible
when additional conductivities due to non-coherent tunneling and
antenna do not compensate the NDC due to coherent interlayer
tunneling, \textit{cf.} Eq.~(\ref{SigmaCond})
\begin{equation}
  \sigma_1 + \sigma^R_{A1}(\omega_0) < \sigma_0/8
. \label{CondNDC}
\end{equation}

Again we consider Eq.~(\ref{equ}), in which we explicitly show the
antenna impedance. As it was mentioned above, on main frequency
the ``friction'' term vanishes, and the equation defining the
frequency of the voltage oscillation reads
$$
  \sigma_0 \tau_p c_s \dot v +
  \frac{i}{S}\Im\left[Z_A^{-1}\right] v = 0
.
$$
Since specific tunneling capacity $c_s$ does not change
dramatically with $v_0$ changes, all conclusions remain valid for
the case of boundaries of large NDC region either.

The oscillation frequency can be presented in the form
\begin{equation}\label{fre}
  \omega_0 = \frac{1}{\tau_p c_s s} \Im \frac{R_0}{Z_A}
.
\end{equation}
Here $R_0$ is resistance of the sample.

Note also, that from equation (\ref{j1pro}) for the
tunneling current it follows that the oscillation frequency can
not be large compared to the momentum scattering rate $1/\tau_p$.
This becomes clear, in particular, from Eq.~(\ref{eqV1pV2w})
demonstrating that the differential conductivity becomes positive
at high frequencies, $\omega > \tau_p^{-1}$.

According to Eq.~(\ref{fre}) oscillation frequency depends on the
input antenna impedance $Z_A$. Dependence $Z_A(\omega)$ can be
very diverse for different antenna configurations, so to
illustrate possibility of the proposed oscillatory regime we
consider the most simple case of the dipole antenna consisting of
thin straight wire of length $l_A$ and radius $r$. This antenna
can be described by the long line theory \cite{Anteny}. Then
antenna wave impedance in the SI units reads $\rho = \frac{1}{\pi
c \varepsilon_0} (\ln \frac{2l_A}{r}-1) [\Omega] = 120 (\ln
\frac{2l_A}{r}-1) [\Omega]$, where $k=2\pi l_A/\lambda$, where
$\lambda$ is the wave length. For the sending-end impedance this
theory gives:
\begin{equation}
  Z_A=\frac{R_{A} \rho^2}{R_{A}^2+\rho^2 \sin^2 k l_a}-
  i\frac{\rho^3 \sin 2k l_A}{2(R_{A}^2+\rho^2 \sin^2 k l_A)}
. \label{Zin}
\end{equation}
Here, the dipole resistance $R_{A}$ is a sum of the radiation
resistance related to the antinode of the current, and of the
ohmic resistance of the antenna's material, the latter is assumed
to be much smaller than the former one. The radiation resistance
is given by expression $R_{\Sigma} = P/I^2 = 30 (2 (\gamma+\ln 2 k
l_A - \mathop{\rm Ci} 2 k l_A)+ (\gamma+\ln k l_A +\mathop{\rm
Ci}4 k l_A-2 \mathop{\rm Ci} 2 k l_A)\cos 2 k l_A + (\mathop{\rm
Si} 4 k l_A - 2 \mathop{\rm Ci}2kl_A)\sin 2 k l_A)[\Omega]$ (here
$P$ is radiation power, $\gamma$ is Euler constant, $\mathop{\rm
Ci}$ and $\mathop{\rm Si}$ are integral cosine and sine,
respectively).

\begin{figure}[!ht]
  \vskip 0mm
  \centerline{
    \psfig{figure=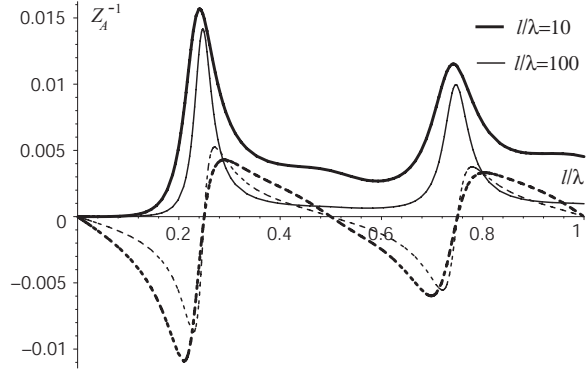,height=5cm
    ,angle=270}
  }
  \caption{
    Dependence of $Z_A^{-1}$ on ratio $l_A/\lambda$ for simple case
    of dipole antenna. Solid and dashed lines correspond
    to active resistance and reactive impedance, respectively.
  }
\label{figZ}
\end{figure}

According to Eq. (\ref{Zin}) the imaginary part of the antenna
impedance is oscillatory function of frequency and on the length
of the antenna arms $l_A$. The dependence of real and imaginary
parts of $Z_A^{-1}$ on $l_A$ is presented in Fig.~\ref{figZ} for
several sets of antenna parameters. When the ratio $l_A/\lambda$
is in the range from 0.5 to 1, $\Im Z_A$ is positive and the
oscillatory solution is possible.

To estimate resistivity of the sample, we use conductivity data
for NbSe$_3$ adduced in Ref.\onlinecite{Latyshev} for NbSe$_3$,
$\sigma_0=0.1 m\Omega\,cm$, and assume the sample dimensions to be
$10 \mu m \times 10 \mu m \times 300$\AA$ $ with the smallest
length in the current direction. Then we obtain $R_0 = 30 \Omega$.
Typical values of $\Re Z_A$ for the most interesting case of
antenna with relatively large wave impedance which corresponds to
high Q-value, are of the order of several $k \Omega$ (see
Fig.~\ref{figZ}), hence, the antenna impedance is larger then the
sample resistance ($R_0 \Im Z_A^{-1} < 1$). So conductivities
$\sigma^I_{An}$ and $\sigma^R_{An}$ are small, as it was supposed
in previous sections.

For practical applications the higher frequencies are more
interesting. We see several possible ways to increase the
oscillation frequency. The first one is to increase $R_0$ by
increasing the ratio $sN/S$, either by changing a sample geometry
or by increasing the number of layers. Implementation of this idea
is limited by the condition for the NDC region to exist
(\ref{CondNDC}) due to an increase of relative dissipation by
antenna caused by radiation as $\sigma^R_{A1} \propto sN/S$.
Another way is to maximize $\Im Z_{A}^{-1}$ in order to optimize
the ratio of the antenna arm length and the wavelength, or to use
more complicated antenna, than the simple dipole one. In this case
the limitations are imposed by real part of antenna impedance
related to radiation power $R_\Sigma$ -- practically interesting
case of larger radiation leads to larger $R_\Sigma$ while $\Im
Z_{A}^{-1} < R_\Sigma^{-1}/2$.

In previous sections we assumed that the current density does not
depend on the in-plane coordinate. In addition to the
non-uniformity related to fluctuations of charge density
considered in Sec.~\ref{secStab}, there is another non-uniformity
which is related to the skin effect. So, strictly speaking, our
results are applicable provided the sample width is smaller, than
the skin-effect length, $c/ \sqrt{2 \pi (\sigma_1 +
\sigma_0/(1+v_0^2)) \omega_0}$. This condition can be easily
satisfied in the tera-Herz region.

In conclusion, we have studied the problem of the current flow
through the sample with narrow energy band when the average
voltage corresponds to the NDC region. We have found analytically
the stable uniform oscillating solution describing coherent
voltage oscillations in the sample. The amplitude and the
frequency of the oscillations are determined by dimensions,
geometry, and material properties of the sample, and depend on the
antenna properties of the system. The maximum value of the
oscillation frequency is limited by the momentum relaxation rate.

\section*{ACKNOWLEDGMENTS}

We are grateful to V.A.~Volkov for helpful comments and to
V.Ya.~Aleshkin for useful information. In part the work was
supported by Russian Foundation for Basic Research (RFBR), by
INTAS, and by CRDF. A part of these research was performed in the
frame of the CNRS-RAS-RFBR Associated European Laboratory
"Physical properties of coherent electronic states in condensed
matter" between CRTBT and IRE RAS.


\begin{thebibliography}{99}
\bibitem{TMD}
R. L. Withers and J. A. Wilson, J. Phys. C \textbf{19}, 4809
(1986).
\bibitem{Sloistost}
  R.\,Kleiner and P.\,M\"uller, Phys. Rev. B \textbf{49}, 1327 (1994).
  R.\,Kleiner, P.\,M\"uller, H.\,Kohlstedt, et al., Phys. Rev. B \textbf{50}, 3942 (1994).
\bibitem{Anisotropy}
  N.P.\,Ong and J.W.\,Brill, Phys. Rev. B \textbf{18}, 5265 (1978);
  Yu.I.\,Latyshev, P.\,Monceau, O.Laborde, et al., J. Phys. IV \textbf{9}, 165 (1999).
\bibitem{Wacker} Andreas Wacker "Semiconductor superlattices: a model system for
nonlinear transport", Physics Reports \textbf{357}, pp. 1-111
(2002).
\bibitem{EsakiTsu} L.\,Esaki and R.\,Tsu, IBM J. Res. Dev., vol. 14, pp. 61-65 (1970);
  L.\,Esaki and R.\,Tsu, Appl. Phys. Lett,  {\bf 19}, 246 (1971).
\bibitem{Artemenko}  S.N.\,Artemenko and A.F.\,Volkov, Fiz. Tverd. Tela (Leningrad) 23,
2153 (1981); Sov. Phys. Solid State \textbf{23}, 1257 (1981)
\bibitem{Latyshev} Yu.I.\,Latyshev, A.A.\,Sinchenko, L.N.\,Bulaevskii, V.N.\,Pavlenko,
P.\,Monceau, JETP Letters, {\bf 75}, 93 (2002)
\bibitem{TauP}
M.I.\,Flik, Z.M.\,Zhang, K.E.\,Goodson, M.P.\,Siegal and Julia
M.\,Phillips, Phys. Rev. B 46, 5606-5614 (1992); D. A. Bonn,
Ruixing Liang, T. M. Riseman, D. J. Baar, D. C. Morgan, Kuan
Zhang, P. Dosanjh, T. L. Duty, A. MacFarlane, G. D. Morris, J. H.
Brewer, W. N. Hardy, C. Kallin and A. J. Berlinsky, Phys. Rev. B
47, 11314-11328 (1993); F. Gao, J. W. Kruse, C. E. Platt, M. Feng,
and M. V. Klein, Appl. Phys. Lett., 63, pp. 2274-2276 (1993).
\bibitem{ACField}
 H.\,Kroemer, cond-mat/0007482 (2000);
  cond-mat/0009311 (2000);
 Yu. A. Romanov, V.P. Bovin, and L. K. Orlov, \\
 Fiz. Tekh. Poluprovodn. \textbf{12}, 1665 (1978) [Sov. Phys. Semicond. \textbf{12}, 987 (1978)]
\bibitem{RomanovImpulsy} Yu.A.\,Romanov and Yu.Yu.\,Romanova, Semiconductors, \textbf{39}, 147 (2005).
\bibitem{RomanovZona}  Yu.A.\,Romanov and Yu.Yu.\,Romanova, Physics of the Solid State, \textbf{46}, 164 (2004).
\bibitem{Landau}E.M. Lifshits and L.P. Pitaevskii, \textit{Physical
kinetics}, (Pergamon Press, London, 1981).
\bibitem{GreenFunction} I.O.\,Kulik and I.K.\,Yanson, Josephson Effect in Superconducting Tunnel Structures, Nauka, Moscow, (1970).
\bibitem{Anteny}   Sazonov D.M., Antennas and Microwaves Devices, High School(USSR), 1988.-432 p.,
  English translation: Microwave Circuits and Antennas, Mir Publishers, Moscow, 1990, 504 p.

\end{thebibliography}
\end{document}